# Site Authorization Service (SAZ)

D Skow
*FNAL, Batavia, IL 60510, USA*

I Mandrichenko
*FNAL, Batavia, IL 60510, USA*

V Sekhri
*FNAL, Batavia, IL 60510, USA*

## 1.ABSTRACT

In this paper we present a methodology to provide an additional level of centralized control for the grid resources. This centralized control is applied to site-wide distribution of various grids and thus provides an upper hand in the maintenance.

## 2.INTRODUCTION

The Site Authorization Service (SAZ) is a service that helps in controlling the user access to any grid resource. With SAZ, Grid resource access gets a two level security check. The first level is that of Local Resource provider and the second level is of SAZ itself. In other words an access to any grid resource on a site has to be authenticated and authorized by both the local mechanism (usually grid-mapfile) and the SAZ. This technique allows centralized control of site wide grids.

## 3.COMPONENTS

SAZ is composed of four components viz. SAZ Server (SAZS), SAZ Database (SAZDB), SAZ User Interface (SAZUI), SAZ Database Admin Interface (SAZDBAI) is shown in Figure 1.

### 3.1. SAZ Server (SAZS)

The SAZ Server is the component that does authentication and authorization of the client who talks to it. The authentication is done using standard token sharing protocol of GSS while authorization is done by simply looking for the client's DN in the database. Some of the tokens are used during the handshake are the part of the client proxy that the client generated before talking to the server. This means that all the clients who talk to the server have to have a proxy first. The server fetches the DN from the proxy and looks it in the database for authorization. If the DN of the client is present in the database, the server returns a successful response, otherwise a failure response. Our SAZ server is written in Java and uses standard libraries of GSS provided by Java Cog Kit. It uses standard JDBC to talk to SAZDB.

Further, SAZ Server is a multi-threaded state-less server. No state of the client who talks to the server is ever remembered. Finally, being multi-threaded allows simultaneous connection to the server.

### 3.2. SAZ Database (SAZDB)

SAZDB stores all the DN's of the clients who are allowed to access any grid resource in the site. The database allows only read access to the SAZ Server while it allows all the privileges to SAZDB Admin Interface. For the sake of simplicity, right now, only DN's are stored in the database. For bookkeeping in future, we will have more information of the client who is using the site resources. The SAZDB is in MySql and does not necessarily have to be on the same machine where the SAZ Server is installed. For better performance it is recommended to have SAZS and SAZDB on different high-end machines.

### 3.3. SAZ User Interface (SAZUI)

These are the clients, which can actually talk to the SAZ Server. The SAZ UI has two different implementations i.e. one in Java and one in C. The Java implementation is mainly for testing purposes once the SAZ Server is installed as it can be tested on any platform. The C implementation on the other hand can be used only on Linux Operating Systems. The reason of having a C implementation is because the globus gatekeeper in also in C and if we want to hook these two together, it is better to use the same language.

These UI's first perform the standard GSS handshake with user proxy and then delegates the user proxy to the SAZ Server. The delegation procedure is completely optional. The SAZ server will work normally even if it just performs the handshake as the DN of the user can be fetched from the handshake if proxies are used.

### 3.4. SAZ Database Admin Interface (SAZDBAI)

This module allows the SAZ admin to do all the operations on the SAZ DB. The Site or SAZ admins can decide which user is allowed or disallowed in the site and can do the corresponding operation on SAZDB using SAZDBAI.

Figure 1: SAZ Architecture

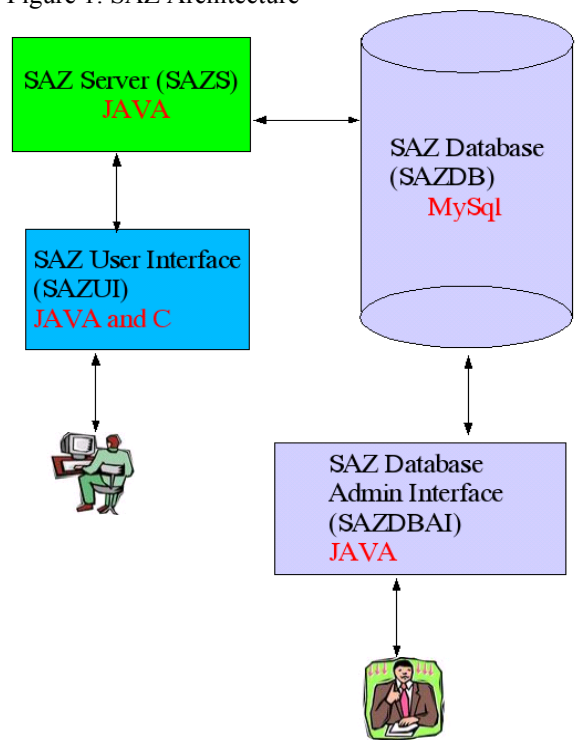

## 4.OPERATIONS

The SAZ User Interface in C is combined with the globus gatekeeper. This is achieved as the new beta release of VDT provides functionality in the gatekeeper to perform callouts during job submission. In other words, the gatekeeper can load the SAZUI module and call it on every job that is submitted. The overall operations consist of the following steps and are shown in Figure 2.

### 4.1. Proxy Generation

Any user who wants to submit a job on any grid on the site has to generate a proxy for him/her using standard proxy generation commands (grid-proxy-init). This makes use of the user certificate that he/she got from any Certificate Authority.

### 4.2. Job Submission

The second step is to just submit the job to any gatekeeper using the standard job submission commands (globus-job-run). At this point the user does not know if he/she is allowed or not to use the grid.

### 4.3. SAZ Authentication

The gatekeeper that has been configured to use SAZUI plugin, will call one of the methods in the plugin. It will also pass the user credentials to this method. The SAZUI opens a socket connection to the SAZ Server and the two will use these user credentials to perform the handshake. This way the user who submitted the job is authenticated to the SAZ Server and a security context is established. Using this security context the SAZUI now follows the SAZ protocol to talk further with the SAZ server. This way the two-way traffic is guaranteed to be secure.

### 4.4. SAZ Authorization

The job of the SAZ server is to first fetch the user DN from the tokens used for handshake or from the credentials that may or may not have been delegated to SAZ server. The main requirement for SAZ to run properly is that the client's credentials have to be delegated to the service (like gatekeeper), from which it can be passed down to SAZUI and finally it can use the credentials for handshake. It is not necessary, that the same credentials be delegated to SAZ server also.

The SAZ server now checks to see if the client is talking with the right SAZ protocol. If the SAZ server is unable to understand the protocol used by the client to communicate, it simply rejects the user credentials and closes the socket. The SAZUI on the client side returns a SAZ failure to the gatekeeper. The job submitted terminates with an authentication error message.

If SAZ server understands the protocol used by the client, it checks for user DN in the SAZDB. If the SAZDB contains the user DN, the SAZ server returns with a success, otherwise it returns with failure and closes the socket. Either success or failure is returned back to the gatekeeper.

### 4.5. Gatekeeper Authorization

On success returned by the SAZUI the gatekeeper now checks the user DN in its local grid-mapfile . This way, another level of authorization is added. Only if both levels of authorization succeed does the get finally submitted.

This ordering of authorization implies that SAZ permission to disallow a user to access the grid will override any allow permission from the local grid-mapfile. On the other hand SAZ permission to allow a user will not override the local disallow permission. Thus to access the resource, the client has to be approved by both level of authorization.

Figure 2: SAZ Steps

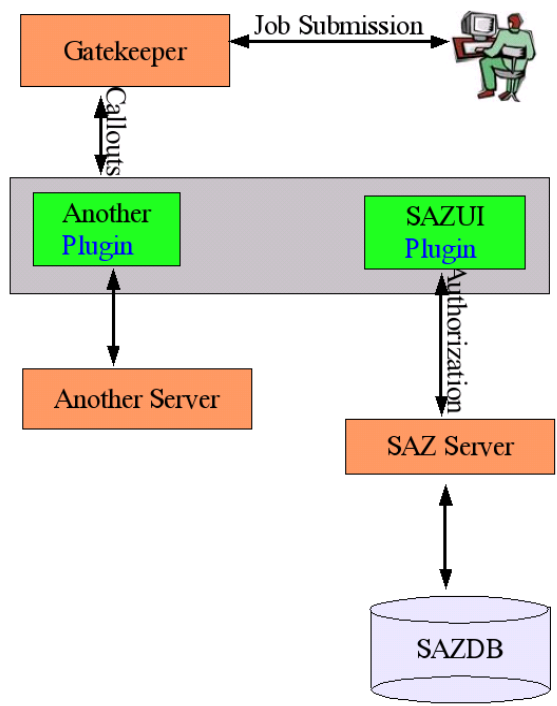

The same protocol can also be used for different types of servers. If a different operation is required performed then the only change will be the operation name. For example we already have another server that allows or denies the user at a particular time. Thus if the user is not allowed at a certain time period, he/she gets back "YES/NO" response respectively. The only change required to talk to this server is to send "TIME" as the operation name instead of "SAZ". This server helps the local admin to control the time slot for allowing and disallowing users. So the same protocol can thus be followed for different types of authorization services.

Figure 3: SAZ Protocol

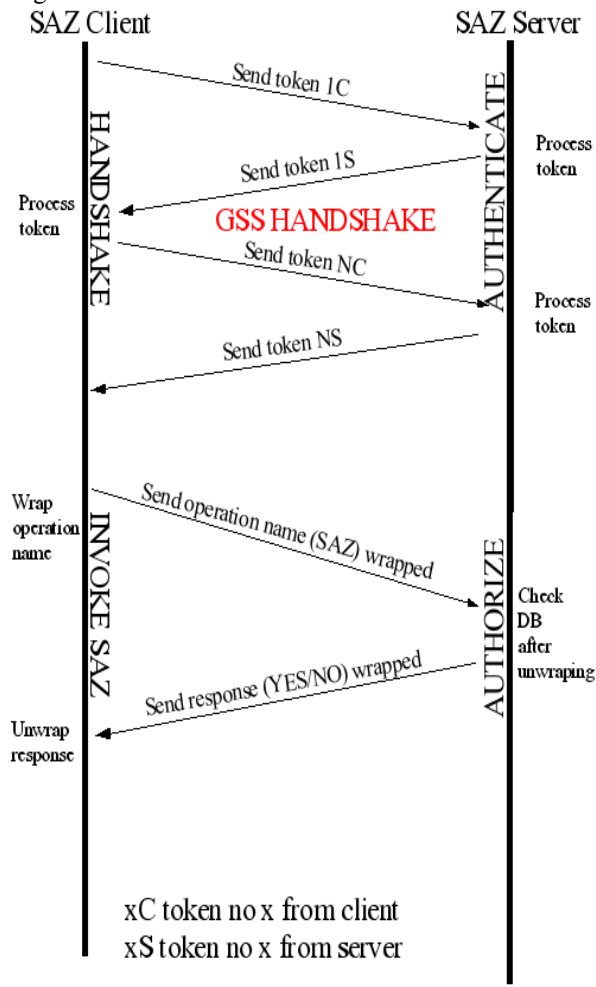

## 5.SAZ PROTOCOL

SAZ follow a very simple and generic protocol. The protocol is designed keeping in mind that the server will be a stateless server and all that the server needs are the credentials. The same protocol can be used for any number of different servers that can be called during job submission.

The first and foremost step is the handshake to the server. This is performed by sending and receiving tokens, which are used for establishing security context. The SAZ server uses standard GSS for performing the handshake and thus uses credentials for authentication. The complete protocol is depicted in Figure 3.

After the client is authenticated, both the server and the client have a security context, which they can use for further communication. This guarantees that no message will ever be tempered with. The client uses this security context to send the operation name that has to be done at the server side. In this case it is "SAZ". The server uses its own security context to see if the operation name is SAZ or not. Then the server searches the database for the user DN to authorize the client. If everything is in place, the server responds with "YES/NO" message and closes the socket. The client depending upon the message received from the server, returns success/failure to the gatekeeper. The gatekeeper submits the job or throws an authentication error message depending on success or failure respectively.

## 6.SAZ STATUS

A Prototype has been tested with EDG gatekeeper. It is now being installed on CMS development testbed and soon on CDF/DO grids. Further specification for site authorization callout has been agreed between Globus, EDG, Virginia Tech, and FNAL teams  and will come with next update of Globus 2.

## 7.FUTURE WORK

Right now SAZ is working perfectly fine with EDG Gatekeeper. The main goal is to get it working with the standard callout in globus gatekeeper. Further, SAZDB may have to store more information other than just DN. This helps in bookkeeping about the users who are using the grid resources. Finally, SAZ may have to be moved into a web service (GTK 3) in Fall 2003.

## 8.REFERENCES

http://www-unix.globus.org/cog/java/
http://www.globus.org/gt2.4/admin/guide-user.html
http://www.dutchgrid.nl/DataGrid/wp4/lcas/edg-lcas-1.0.3/